\documentclass[epsfig,indentfirst,bm,12pt]{article}
\usepackage{CJK}
\usepackage{graphicx,amsmath,dcolumn,bm}
\begin{document}
\title{\Large{\bf{Hadron formation in semi-inclusive deep inelastic lepton-nucleus scattering }}
\thanks {Supported partially by  National Natural Science Foundation of China (11075044) and
Natural Science Foundation of Hebei Province (A2008000137).}}
\begin{CJK*}{GBK}{song}
\author{Li-Hua Song $^{1,3}$
 Na Liu $^{1,4}$
 Chun-Gui Duan $^{1,2}$
\footnote{\tt{ E-mail:duancg$@$mail.hebtu.edu.cn}}}

\date{}

\maketitle

\end{CJK*}

\noindent {\small 1.Department of Physics, Hebei Normal
               University, Shijiazhuang 050024, China}\\
{\small 2.Hebei Advanced Thin Films Laboratory, Shijiazhuang 050024, China}\\
{\small 3.College of Science,  Hebei United University, Tangshan
063009, China}\\
{\small 4.College of Mathematics and Physics, Shijiazhuang
University of Economics, Shijiazhuang 050031, China}

\baselineskip 9mm
\begin{abstract}

Hadron production in lepton-nucleus deep inelastic scattering is
studied in a model including quark energy loss and nuclear
absorption. The leading-order computations for hadron multiplicity
ratios are presented and compared with the selected HERMES
experimental data with the quark hadronization occurring inside the
nucleus by means of the hadron formation time. It is shown that with
increase of the energy fraction carried by the hadron,  the nuclear
suppression on hadron multiplicity ratio  from nuclear absorption
gets bigger. It is found that when hadronization occurs inside the
nucleus, the nuclear absorption is the dominant mechanism causing a
reduction of the hadron yield. The atomic mass dependence of hadron
attenuation for quark hadronization starting inside the nucleus is
confirmed theoretically and experimentally to be proportional to $
A^{1/3}$.



\end{abstract}

\maketitle
\newpage
\vskip 0.5cm

\section*{1. Introduction}

The semi-inclusive deep inelastic scattering of lepton on nuclear
target has been one of the most active frontiers in nuclear physics
and particle physics over the past three decades. After the
pioneering measurement of hadronization in the nuclear medium at
SLAC$^{[1]}$, more precise data were reported by the
HERMES$^{[2-6]}$ and CLAS$^{[7]}$  Collaborations. It is desirable
that the quantitative information on the quark propagation and
hadronization in the nuclear medium can be obtained by means of  the
observed hadron distributions and multiplicities from various
nuclei. Furthermore, the quantitative information can provide
enlightenments and references for  the study of quark-gluon plasma
and its space-time evolution in ultra-relativistic heavy-ion
collisions.

In the semi-inclusive deep inelastic scattering on nuclei, a virtual
photon from the incident lepton is absorbed by a quark within a
nucleus, the highly virtual colored quark traverses over some
distance through the cold nuclear medium, evolves subsequently into
an observed hadron. There is a key physical quantity, i.e., the
characteristic time of quark propagation, or so-called hadron
formation time. The hadron formation time is referred in detail to
the time between the moment that the quark is struck by the virtual
photon and the moment that the prehadron (the predecessor of the
final hadron) is formed. If the hadron formation time is short
enough, the prehadron is formed inside the target nucleus. In
addition of the struck quark energy loss due to multiple
interactions with the surrounding nuclear medium and gluon
radiation, the prehadron can interact via the relevant hadronic
interaction cross section, causing a further reduction of the
observed hadron yield. Thereupon, when hadronization takes place
inside the nucleus, semi-inclusive deep inelastic lepton-nucleus
collisions can provide the important information on the space-time
development of the hadronization process in the nuclear medium. It
is worth emphasizing that hadronization in the nuclear medium are
intrinsically non-perturbative QCD processes. The perturbative
calculation cannot be applied to evaluate the underlying
interactions in the QCD framework.

Two classes of theoretical phenomenological models, the
absorption-type models$^{[8-10]}$ and the parton energy loss
models$^{[11-13]}$,  were proposed to investigate the experimental
data from the semi-inclusive deep inelastic scattering of lepton on
the nucleus$^{[14]}$. However, the two extreme mechanisms do not
give the best description  of the experimental measurements. In
addition, whatever the physical mechanism, the atomic mass
dependence for the attenuation of hadron production will be an
important ingredient. Nonetheless, the recent researches$^{[15,16]}$
indicate that atomic mass dependence of hadron production is far
from being expected by the absorption and parton energy loss models.
Therefore, it is expected that the atomic mass dependence of hadron
production would be allowed to discriminate between the two
different mechanisms.

In our preceding article$^{[17]}$, by means of the  hadron formation
time, the relevant data with quark hadronization occurring outside
the nucleus are picked out from HERMES experimental results$^{[5]}$
on the one-dimensional dependence of the multiplicity ratio as a
function of the energy fraction $z$.   We have calculated the
nuclear modifications of hadron production in semi-inclusive deep
inelastic scattering  in a parton energy loss model. The  energy
loss per unit length is obtained for an outgoing quark by the global
fit of the selected experimental data. In this paper we employ the
so called two dimensional data from recent HERMES experiment on the
multiplicity ratio for the production of positively and negatively
charged pions and kaons$^{[6]}$. The experimental data with quark
hadronization occurring inside the nucleus are selected by means of
the  hadron formation time. The quark propagation and hadronization
are studied in the nuclear medium.

The remainder of the paper is organized as follows. In Section 2,
the brief formalism for the hadron  multiplicity in semi-inclusive
deep inelastic scattering on the nucleus. In Section 3, the
numerical computations for the multiplicity ratio are presented and
compared with experimental data. Finally, a summary is presented.

\section*{2. The theoretical framework }

At leading order(LO) in perturbative QCD, the hadron multiplicity
can be obtained from normalizing the semi-inclusive deep inelastic
lepton  nucleus scattering yield $N^{h}_{A}$ to the deep inelastic
scattering yield $N^{DIS}_{A}$,
\begin{equation}
\frac{1}{N^{DIS}_{A}}\frac{dN^{h}_{A}}{dzd
\nu}=\frac{1}{\sigma^{lA}}\int
dx\sum_{f}e^{2}_{f}q^{A}_{f}(x,Q^{2})\frac{d\sigma^{lq}}{dxd\nu}D^{A}_{f|
h}(z,Q^{2}),
\end{equation}
\begin{equation}
\sigma^{lA}=\int
dx\sum_{f}e^{2}_{f}q^{A}_{f}(x,Q^{2})\frac{d\sigma^{lq}}{dxd\nu},
\end{equation}
\begin{equation}
\frac{d\sigma^{lq}}{dxd\nu}=Mx\frac{4\pi\alpha_{s}^{2}}{Q^{4}}[1+(1-y)^{2}].
\end{equation}
In the above equation, $\nu$ is the virtual photon energy, $e_f$ is
the charge of the quark with flavor $f$, $q^{A}_{f}(x,Q^{2})$ is the
nuclear quark distribution function with Bjorken variable $x$ and
photon virtuality $Q^{2}$, ${d\sigma^{lq}}/{dxd\nu}$ is the
differential cross section for lepton-quark scattering at leading
order,  $D^{A}_{f| h}(z,Q^{2})$ is the nuclear modified
fragmentation function of a quark of flavour $f$ into a hadron $h$,
and $ \alpha _{s}$ and $y$ are the fine structure constant and the
fraction of the incident lepton energy transferred to the target,
respectively.

After a quark within a nucleus interacts with a virtual photon from
the incident lepton,  the struck quark can lose its energy owing to
multiple scattering and gluon radiation while propagating through
the nucleus. The quark energy fragmenting into a hadron shifts from
$E_{q}=\nu$ to $E'_{q}=\nu-\Delta E$, which results in a rescaling
of the energy fraction of the produced hadron:
\begin{equation}
z=\frac{E_{h}}{\nu}  \longrightarrow  z'=\frac{E_{h}}{\nu-\Delta E},
\end{equation}
where $E_{h}$ and $\Delta E$ are, respectively, the measured hadron
energy and the quark energy loss in the nuclear medium. Thus, the
fragmentation function in the nuclear medium is assumed to
be$^{[17]}$
\begin{equation}
D^{A}_{f|h}(z,Q^{2})=D_{f|h}(z',Q^{2}),
\end{equation}
where $D_{f|h}$ is the standard (vacuum) fragmentation function of a
quark of flavour $f$ into a hadron $h$.

In order to further consider the hadronization in the nuclear
medium, it is assumed that the quack struck by the virtual photon at
a longitudinal position $y$ fragments into a prehadronic state at
the point $y'$ and soon after the final hadron is created. The
prehadronic state and the final hadron propagate through nuclear
matter and interact with the surrounding nucleons in the target
nucleus, causing a further reduction of the observed hadron yield.
In the case of treating  the  prehadron and the hadron as a single
object created at $y'$,  Bialas and Chmaj$^{[18]}$ provided the
probability $P_{h}(y',y)$ that the  virtual colored quark fragments
into a hadron at a distance $y'-y$,
\begin{equation}
P_{h}(y',y)=1-e^{-(y'-y)/t},
\end{equation}
where $t$ is the hadron formation time. The survival probability
$S_{A}(b,y)^{A-1}$ of this hadron can be obtained from
\begin{equation}
S_{A}(b,y)=1-\sigma_{h}\int_{y}^{\infty}dy'P_{h}(y',y)\rho_{A}(b,y').
\end{equation}
where $\sigma_h$  is the hadron-nucleon inelastic cross section, and
$\rho_{A}(b,y')$ is the nuclear density profile normalized to unity
as a function of impact parameter $b$ and longitudinal coordinate
$y'$. Furthermore,  the nuclear absorption factor $N_{A}(z,\nu)$,
which is defined as the probability that neither the prehadron nor
hadron have interacted with a nucleon, is given as,
\begin{equation}
N_{A}(z,\nu)=\int
d^{2}b\int_{-\infty}^{\infty}dy\rho_{A}(b,y)S_{A}(b,y)^{A-1}.
\end{equation}

With combining the influence of nuclear absorption of the final
hadron and quark energy loss on the hadron production occurring
inside the nucleus, the hadron multiplicity can be presented as
\begin{equation}
\frac{1}{N^{DIS}_{A}}\frac{dN^{h}_{A}}{dzd
\nu}=\frac{1}{\sigma^{lA}}\int
dx\sum_{f}e^{2}_{f}q^{A}_{f}(x,Q^{2})\frac{d\sigma^{lq}}{dxd\nu}D_{f|h}(z',Q^{2})N_{A}(z,\nu).
\end{equation}

\section*{3. Results and discussion}

The HERMES Collaboration  recently reported the two-dimensional data
on the multiplicity ratios for positively and negatively charged
pions and kaons produced on neon, krypton, and xenon targets
relative to deuterium, in three z slices as a function of $\nu$, and
in three $\nu$ slices as a function of $z$.  We pick out the
experimental data with quark hadronization occurring inside the
nucleus by means of the hadron formation time
$t=z^{0.35}(1-z)\nu/\kappa$ ($\kappa = 1GeV/fm$)$^{[19]}$. More
specifically, if $t$ is less than $L_{A}$ ($L_{A}$ = $3/4R_A$fm),
hadronization occurs inside the nucleus. Otherwise, hadrons are
produced outside the target nucleus. As for the cases with quark
hadronization occurring inside the nucleus, we compute at leading
order  the hadron multiplicity ratios $R^{h}_{A}$,
\begin{equation}
R^{h}_{A}[z]=\int
\frac{1}{N^{DIS}_{A}}\frac{dN^{h}_{A}(\nu,z)}{dzd\nu}d\nu \Bigg/\int
\frac{1}{N^{DIS}_{D}}\frac{dN^{h}_{D}(\nu,z)}{dzd\nu}d\nu.
\end{equation}
In the calculation, the nuclear effects on deuterium target are
ignored.  The CTEQ6L parton density in the proton$^{[20]}$ is used
together with the vacuum fragmentation functions $^{[21]}$.

Concerning the energy loss of the struck quark while propagating
through the nucleus, the linear quark energy loss
parametrization$^{[22,23]}$ was employed, which is written as
\begin{equation}
\Delta E=\alpha L_{A},
\end{equation}
where $\alpha$ is the parameter that can be extracted from
experimental data, and $L_{A}$ is the path length a quark takes in
the target nucleus. As for the nuclear density profile, the
Woods-Saxon distribution was used,
\begin{equation}
\rho(r)=\rho_0/(1+\exp(r-r_A)/a),
\end{equation}
where $a=0.545$, and $r_A=1.12A^{1/3}$fm$^{[24]}$.

In order to determine the  parameter $\alpha$ in quark energy loss
expression and the cross-section $\sigma_{h}$ in nuclear absorption
factor, the calculated hadron multiplicity ratios $R^{h}_{A}$ are
compared with the selected experimental values by using the CERN
subroutine MINUIT $^{[25]}$ and minimizing $\chi^2$,
\begin{equation}
\chi^{2}=\sum_{i}^{m} \Bigg[\frac{R^{h, data}_{A, i}-R^{h, theo}_{A,
i}}{\sigma^{err}_{i}} \Bigg]^2,
\end{equation}
where $R^{h, data}_{A, i}$ and $R^{h, theo}_{A, i}$ indicate
separately the experimental data and theoretical values of the
hadron multiplicity ratio $R^{h}_{A}$, the experimental error is
given by systematic and statistical errors as
$(\sigma^{err}_{i})^2=(\sigma^{syst}_{i})^2+(\sigma^{stat}_{i})^2$.
The uncertainties of the optimum $\alpha$ and $\sigma_{h}$ are
obtained with an increase of $\chi^{2}$ by 4.61 unit from the
minimum $\chi^{2}_{min}$,  which corresponds, in the case of
normally distributed fit parameters, to the $90\%$  covered region
of the total probability distribution.

\begin{table}[t,m,b]
\caption{The values of parameters $\alpha$, $\sigma_{h}$   and
$\chi^{2}/ndf$
 for positively and negatively charged pions and kaons
 extracted from the selected data on $R^{h}_{A}$ for
hadron produced on krypton and xenon nuclei, as  a function of $z$
in $4<\nu<11$ GeV and  $11<\nu<14$ GeV.}
\begin{center}
\begin{tabular}{p{4cm}p{3cm}p{3cm}p{3cm}c}\hline
          Identified hadron  & $\alpha(GeV/fm)$ &$\sigma_{h}(mb)$&$\chi^{2}/ndf $\\
         \hline
         \multicolumn{1}{c}{$\pi^{+}$}  & $0.09\pm0.01$&$29.8\pm0.1$&0.34\\
       \multicolumn{1}{c}{$\pi^{-}$}&$0.06\pm0.01$&$30.0\pm0.2$&0.24\\
        \multicolumn{1}{c}{ $k^{+}$ }      & $0.06\pm0.02$&$18.7\pm0.2$&0.22\\
      \multicolumn{1}{c}{ $k^{-}$ } & $0.14\pm0.04$&$21.8\pm1.1$&0.47\\

\hline
\end{tabular}
\end{center}
\end{table}

Table 1 summarizes $\chi^2$ per number of degrees of freedom
($\chi^2/ndf$), the determined parameters $\alpha$ and  $\sigma_{h}$
 by calculating the hadron multiplicity
ratios $R^{h}_A$ for positively and negatively charged pions and
kaons production on krypton nucleus relative to the deuteron as a
function of $z$ in $4<\nu<11$ GeV  and on xenon nucleus in
$4<\nu<11$ GeV  as well as $11<\nu<14$ GeV from the selected HERMES
experiment data$^{[6]}$. It is shown that the theoretical results
with considering the nuclear absorption  and the nuclear
modification of fragmentation functions owing to quark energy loss
give an excellent description of the experimental data with the
values of $\chi^{2}/ndf $ being far smaller than 1. From Table 1, it
can be seen that the quark energy loss required to describe the
measurements for hadronization occurring inside the nucleus can not
be properly determined for negatively charged pions and  kaons
production because the relative uncertainty $\delta \alpha / \alpha
$ is very large. In detail, $\delta\alpha /\alpha$ are approximately
$17\%$, $33\%$ and $28\%$ for negatively  charged pions and
positively and negatively kaons production, respectively. However,
$\delta \alpha / \alpha \simeq 11\%$ for positively charged pions
production, the energy loss per unit length is pinned down as
$\alpha=0.09$ GeV/fm. Of particular concern is that the value of
$\alpha$ is much smaller than the one from our paper$^{[17]}$. The
reason causing this result is worthy of our further investigating.

It is believed that the  colored quark created at the virtual
photon-quark interaction point can lose its energy with experiencing
multiple interactions with the surrounding nuclear medium and
induced gluon radiation. Therefore, the  energy loss per unit length
is put to $\alpha=0.09$ GeV/fm.  The values of $\sigma_{h}$ and
$\chi^{2}/ndf$ extracted from the selected data on $R^{h}_{A}$ are
listed in Table 2 by means of the $\chi^2$ analysis. It can be seen
that the agreement between the theoretical calculations and selected
data is yet good. The fit of  the selected data in the present
analysis leads to the  cross sections of $29.9\pm0.2$ mb,
$26.3\pm0.2$ mb,  $15.5\pm0.1$ mb and $28.7\pm0.8$ mb for positively
and negatively charged pions and kaons.

\begin{table}[t,m,b]
\caption{The values of $\sigma_{h}$ and $\chi^{2}/ndf$ extracted
from the selected data on $R^{h}_{A}$ with $\alpha=0.09$ GeV/fm.}
\begin{center}
\begin{tabular}{p{4cm}p{4cm}p{4cm}c}\hline
          Identified hadron   &$\sigma_{h}(mb)$&$\chi^{2}/ndf $\\
         \hline
         \multicolumn{1}{c}{$\pi^{+}$}  & $29.9\pm0.2$&0.31\\
       \multicolumn{1}{c}{$\pi^{-}$}&$26.3\pm0.2$&0.30\\
        \multicolumn{1}{c}{ $k^{+}$ }      &$15.5\pm0.1$&0.23\\
      \multicolumn{1}{c}{ $k^{-}$ } &$28.7\pm0.8$&0.44\\

\hline
\end{tabular}
\end{center}
\end{table}


\begin{figure}[t,m,b]
\centering
\includegraphics*[width=14cm,height=7cm]{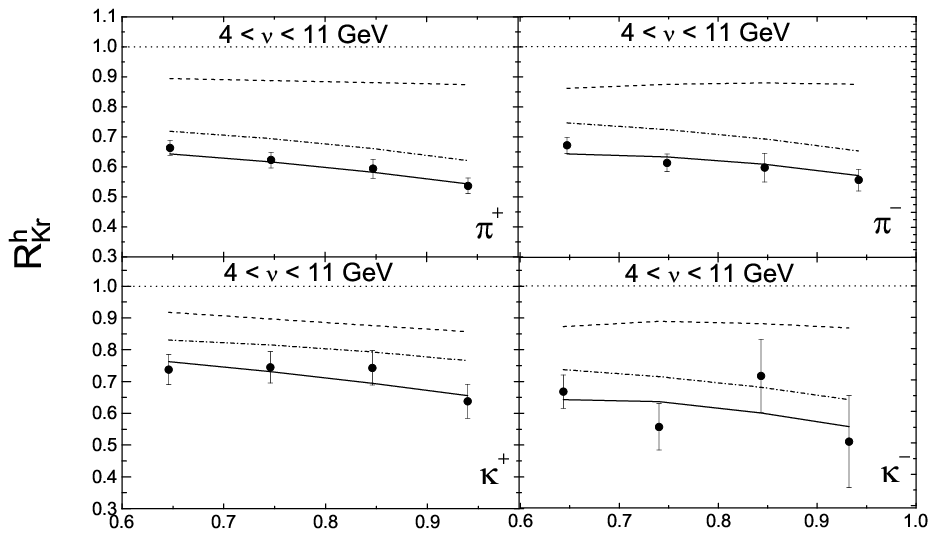}
\includegraphics*[width=14cm,height=7cm]{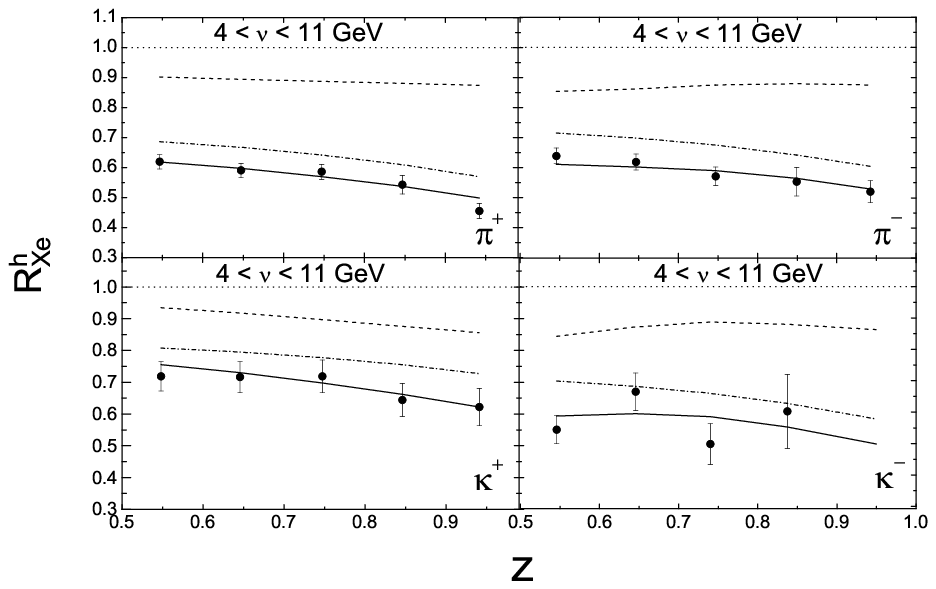}
\includegraphics*[width=14cm,height=7cm]{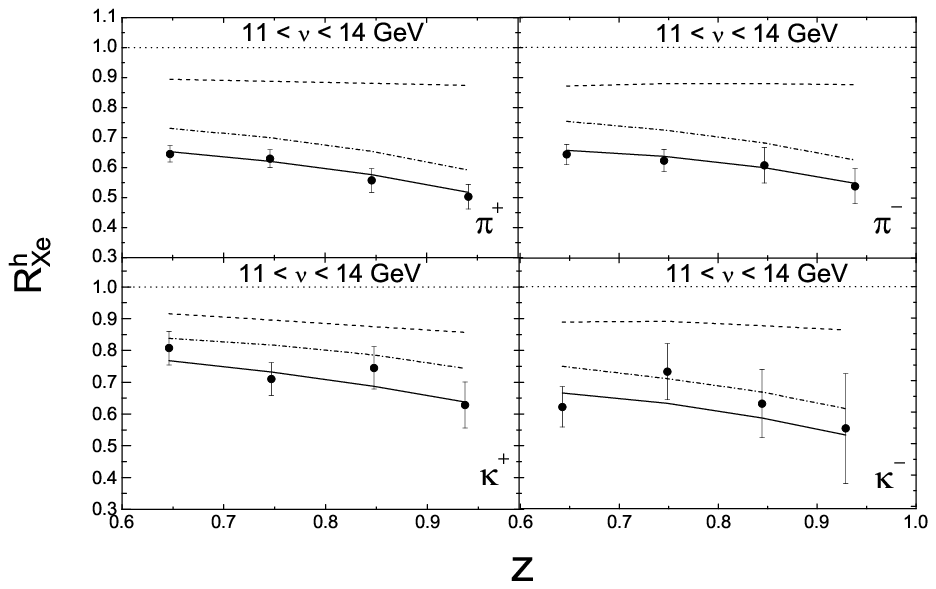}
\vspace{-0.3cm} \caption{The calculated multiplicity ratios
$R^{h}_{A}$ for positively and negatively charged pions and
 kaons production on Kr
 and Xe nuclei with the combination of
nuclear absorption and quark energy loss (solid line), only
considering quark energy loss (dashed line), and only nuclear
absorption (dash dot line). The relative optimal parameters
$\sigma_{h}$ are taken from Table 2 with $\alpha=0.09$ GeV/fm. The
HERMES data$^{[6]}$ on Kr and Xe nuclei are shown with the total
uncertainty (statistical plus systematic, added quadratically).}
\end{figure}


In order to explore respectively the influence of nuclear absorption
and quark energy loss on the  hadron multiplicity ratios $R^{h}_A$,
the theoretical multiplicity ratios are compared with the selected
HERMES experimental data  for positively and negatively charged
pions and  kaons production in Fig.1. The dashed, dash dot, and
solid lines are the results on $R^{h}_A$ by considering the quark
energy loss, nuclear absorption and nuclear absorption plus quark
energy loss, respectively. It is apparent that the computed results
with the combination of nuclear absorption and quark energy loss are
in good agreement with the experimental data. In theory, when the
value of $z$ increases to 1, the nuclear suppression on $R^{h}_A$
from quark energy loss effect is diminished gradually.  However, the
ignoration of nuclear effects on deuterium target gives this feature
on the quark energy loss effect in Fig.1. The nuclear absorption
effect adds further to the nuclear suppression on $R^{h}_A$.  Taking
the example of the hadron multiplicity ratios $R^{h}_A(z)$ in the
region of $0.65\leq z \leq 0.94$ in $4<\nu<11$ GeV for positively
pions production on krypton, the influence of quark energy loss
changes  from $11\%$ to $14\%$ while the impact of nuclear
absorption increases from $28\%$ to $38\%$ with the increase of $z$.
Therefore, it can be concluded that the nuclear absorption becomes
the dominant effect for the selected experimental data with quark
hadronization occurring inside the nucleus.

In order to investigate theoretically the atomic mass dependence of
hadron production in semi-inclusive deep inelastic lepton-nucleus
scattering for  quark hadronization occurring inside the nucleus,
the hadron attenuation $1-R^{h}_{A}$ is presumed in terms of a power
law:
\begin{equation}
1-R^{h}_{A}=cA^{1/3(2/3)}.
\end{equation}
In general, the coefficient $c$  depends on the kinematic variable
$z(\nu)$ and the atomic mass number $A$. The optimal parameter $c$
 in  the power law can be determined by chi-square
minimization, i.e., the $\chi^{2}$ merit function
\begin{equation}
\chi^{2}(c)=\sum_{i}^{m}
\Bigg[\frac{(1-R^{h}_{A})(A_i)-cA^{1/3(2/3)}_i}{\sigma^{err}_{i}}
\Bigg]^2,
\end{equation}
is minimized with respect to the coefficient $c$. Here
$\sigma^{err}_{i}$ is the uncertainty of the theoretical value
$R^{h}_{A}$ due to the cross section $\sigma_h$. We evaluate the
uncertainty of $R^{h}_{A}[z, \sigma_{h}]$ with respect to the
optimized cross-section $\sigma_{h}$ by using Hessian method and
assuming linear error propagation:
\begin{equation}
\delta R^{h}_{A}[z, \sigma_{h}]=\triangle\chi^{2} \frac{\partial
R^{h}_{A}[z, \sigma_{h}]}{\partial \sigma_{h}}\Bigg[
\frac{\partial^2 R^{h}_{A}[z, \sigma_{h}]}{\partial^2 \sigma_{h}}
\Bigg]^{-1/2}.
\end{equation}
In our estimation, $\triangle\chi^{2}=1$ with the confidence level
$P=0.6826$.

We compute the hadron multiplicity ratio $R^{h}_{A}$ for positively
and negatively charged pions and  kaons production on krypton,
stannum, xenon and tungsten nuclei relative to deuterium target in
our model of the quark energy loss plus nuclear absorption. In order
to compare with the relative experimental data, the regions of
kinematics variable $\nu$ and $z$ are limited to being the same as
those from the selected data on krypton and xenon nuclei with quark
hadronization occurring inside the nucleus. In our calculation, the
energy loss per unit length $\alpha=0.09$ GeV/fm, the cross sections
$\sigma_h$ for positively and negatively charged pions and
 kaons are separately taken from Table 2.
The obtained best-fit coefficient $c$ with its uncertainty and
$\chi^2/ndf$ for the fit $1-R^{h}_{A}=cA^{1/3}$ ($cA^{2/3}$)
indicate that the atomic mass dependence of hadron production for
quark hadronization starting inside the nucleus is in good agreement
with  $1- R^h_A \sim A^{1/3}$.

\begin{figure}[t,m,b]
\centering
\includegraphics*[width=14cm,height=7cm]{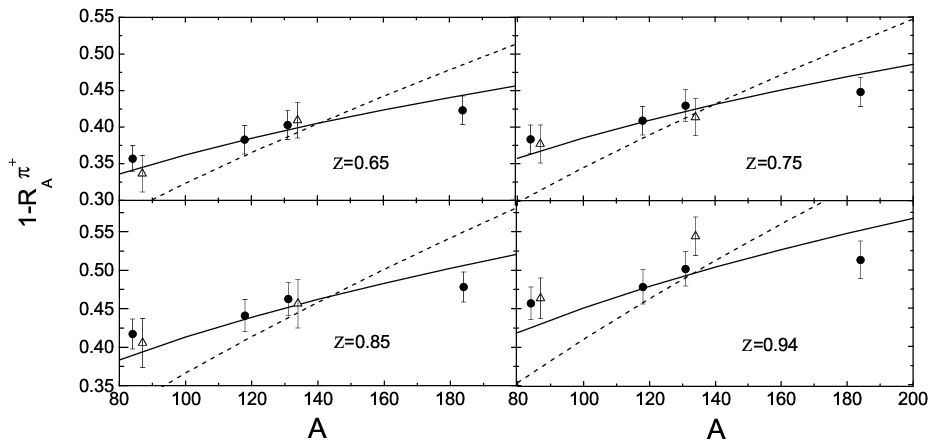}
\includegraphics*[width=14cm,height=7cm]{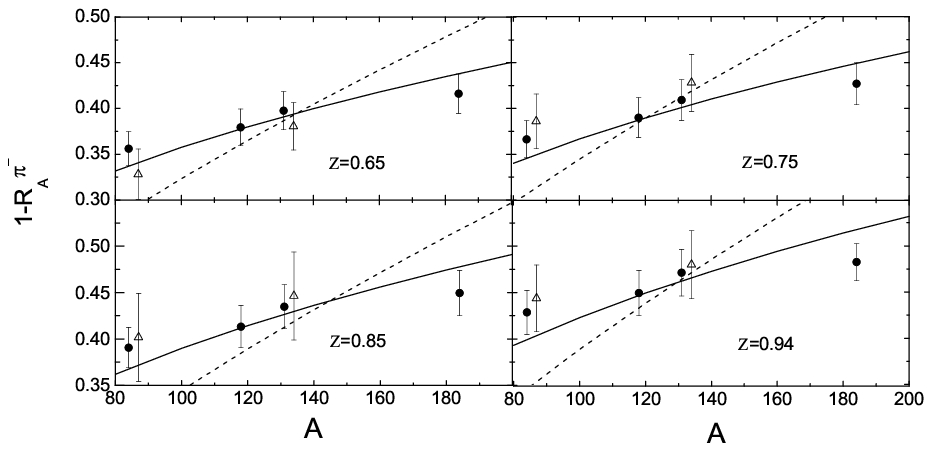}
\vspace{-0.25cm} \caption{The selected HERMES data on
$R^{h}_A$($h=\pi^{+}, \pi^{-}$) for $^{84}Kr$  and
$^{131}Xe$$^{[6]}$ are presented as $1-R^{h}_A$ for various $z$
values as empty triangles, respectively. Our model results for
$1-R^{h}_A$ including values for A = 84, 118, 131, 184 (Kr, Sn, Xe,
W) are shown by solid circles. Note that the experimental points
(empty triangle ) are shifted slightly to the right to avoid overlap
with our model values. The solid lines (dashed lines) are the
calculated results to $1-R^{h}_{A}=cA^{1/3}$ ($cA^{2/3}$).}
\end{figure}

\begin{figure}[t!]
\centering
\includegraphics*[width=14cm,height=7cm]{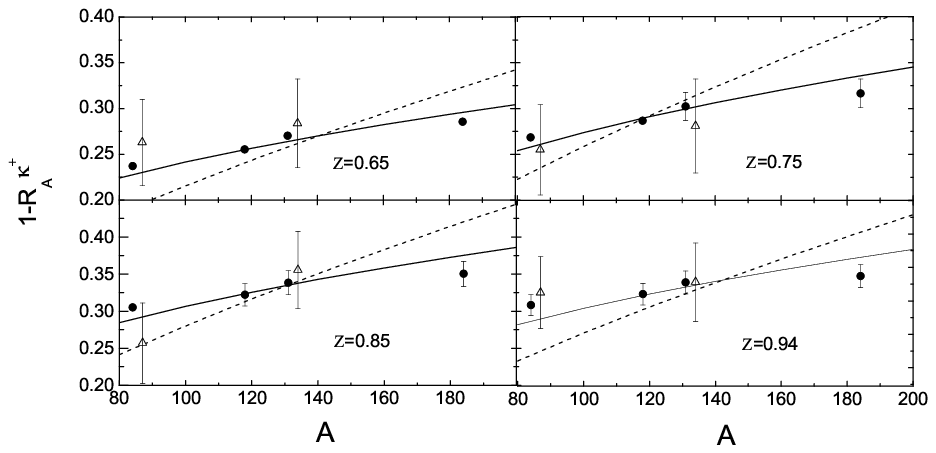}
\includegraphics*[width=14cm,height=7cm]{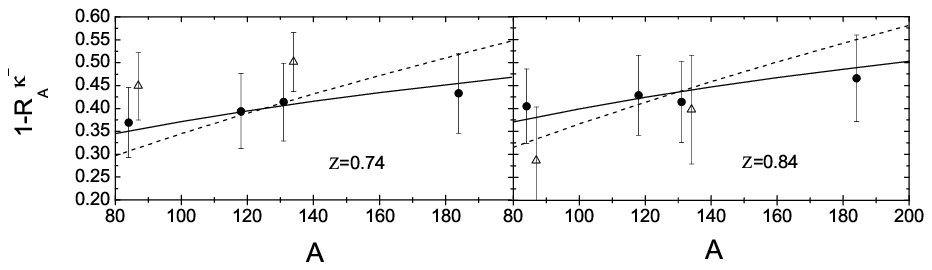}
\vspace{-0.25cm} \caption{The selected HERMES data on
$R^{h}_A$($h=k^{+}, k^{-}$) for $^{84}Kr$  and $^{131}Xe$$^{[6]}$.
The other comments are the same as those in Fig.2.}
\end{figure}


The theoretical results from our model  are compared with the
selected HERMES experimental data  for various hadron production in
Fig.2 and Fig.3.  In the two figures the selected HERMES data on
$R^{h}_A$ for $^{84}Kr$ and $^{131}Xe$$^{[6]}$  are presented as
$1-R^{h}_A$ for the various selected $z$-bins  as empty triangles.
The solid circles show the corresponding results of our model for
four species hadron production on krypton, stannum, xenon and
tungsten nuclei relative to deuterium target. In addition the solid
(dashed) lines are the best fits to $1-R^{h}_{A}=cA^{1/3}$
($cA^{2/3}$) of the theory results. It is worth mentioning that the
experimental points are shifted slightly to the right to avoid
overlap with our model values. It is found from Fig.2 and Fig.3 that
our model is  in support of the power law $1- R^h_A \sim A^{1/3}$.
From what has been discussed above, when hadronization occurs inside
the nucleus, the nuclear absorption is the dominant mechanism which
causes a reduction of the hadron yield. Therefore, we can confirm
the theoretical prediction from the nuclear absorption model on the
power law: $1- R^h_A \sim A^{1/3}$. Overall, the comparison between
the experimental and theoretical results shows they are in
reasonable agreement. The fact demonstrates that the experimental
data is consistent with the $A^{1/3}$ power law for quark
hadronization occurring inside the nucleus.

Moreover, special emphasis is that our conclusion on the power law
is apparently different from the ones from Ref.15 and 16. The
authors of Ref.15 find that contrary to common expectations for
absorption models, the mass number dependence of the hadron
attenuation $1- R^h_A$ is proportional to $A^{2/3}$ in leading
order. In addition, the study on the attenuation of hadron
production by using realistic matter distributions$^{[16]}$ shows
that the mass number dependence for a pure partonic (absorption)
mechanism is more complicated than a simple $A^{2/3}(A^{1/3})$
behavior.

\section*{4. Summary }

Hadron production in lepton-nucleus deep inelastic scattering is
studied in our model including quark energy loss and nuclear
absorption. In the proposed model, the experimental data with the
quarks hadronization occurring inside the nucleus are selected by
means of the hadron formation time. We perform a leading order
phenomenological analysis on the hadron multiplicity ratio, and
compare with the HERMES experimental results for hadron produced on
krypton and xenon nuclei. It is found that the nuclear suppression
on hadron multiplicity ratio from nuclear absorption gets bigger as
the kinematics variable $z$ rises. When hadronization occurs inside
the nucleus, the nuclear absorption is the dominant mechanism
causing a reduction of the hadron yield. The atomic mass dependence
of hadron production for quark hadronization starting inside the
nucleus is theoretically and experimentally in good agreement with
$1- R^h_A \sim A^{1/3}$.

What we should note is that the value of $\alpha$ is much smaller
than that from our previous article. The reason causing this result
is worthy of our further studying. Also, we suggest that future
semi-inclusive deep-inelastic scattering measurements should
increase the amount of target nucleus, and collect data up to heavy
nuclei like Sn and Pb. The precise experimental result will help
unraveling the space-time dynamics of the hadronization process.

\vskip 1cm

\end{document}